\documentstyle[preprint,aps,epsfig]{revtex}
\newif\iftightenlines\tightenlinesfalse
\tightenlines\tightenlinestrue

\begin{document}

\title{
Searching for $\nu_\mu \rightarrow \nu_\tau$ Oscillations with Extragalactic 
Neutrinos}
\author{Sharada Iyer$^1$, Mary Hall Reno$^{2}$ and Ina Sarcevic$^1$}
\address{
$^1$Department of Physics, University of Arizona, Tucson, Arizona
85721\\
$^2$Department of Physics and Astronomy, University of Iowa, Iowa City,
Iowa 52242
}

\maketitle

\begin{abstract}
We propose a novel approach for studying 
$\nu_\mu \rightarrow \nu_\tau$ oscillations with extragalactic 
neutrinos.  Active Galactic Nuclei and Gamma Ray Bursts 
are believed to 
be  sources of ultrahigh energy muon neutrinos.  With distances of 
100 Mpc or more, they provide an unusually long baseline for 
possible detection of 
$\nu_\mu \rightarrow \nu_\tau$ with mixing parameters 
$\Delta m^2$ down to $10^{-17}$eV$^2$, many orders of magnitude 
below the current accelerator experiments.  By solving the coupled 
transport equations, we show that high-energy 
$\nu_\tau$'s, as they propagate through the earth, cascade down in 
energy, 
producing the enhancement of the incoming $\nu_\tau$ flux in the low energy 
region, in 
contrast to the 
high-energy $\nu_\mu$'s, which get absorbed.  
For an AGN quasar model we find the $\nu_\tau$ flux 
to be a factor of $2$ to $2.5$ larger than the incoming flux 
in the energy range 
between $10^2$~GeV and 
$10^4$~GeV, while for a GRB fireball 
model, the enhancement is $10\%$-$27\%$ in 
the same energy range and for zero nadir angle.  
This enhancement decreases with larger nadir 
angle, thus providing a novel way to search for $\nu_\tau$ appearance 
by measuring the angular dependence of the muons.  
To illustrate 
how the cascade effect and the $\nu_\tau$ final flux depend on 
the steepness of the incoming $\nu_\tau$, 
we show the energy and angular distributions 
for several generic cases of the 
incoming tau neutrino flux, $F_\nu^0 \sim E^{-n}$ for $n=1,2$ and $3.6$.  
We show that for the incoming flux that is not too steep, 
the signal for the 
appearance of high-energy 
$\nu_\tau$ is the enhanced production of lower energy 
$\mu$ and their distinctive angular dependence, 
due to the contribution from the $\tau$ decay into $\mu$ just below the 
detector.  

\end{abstract}

\vskip 0.1true in

Recent Super-Kamiokande (SuperK) measurements of the
low atmospheric $\nu_\mu/\nu_e$ ratio and the strong
zenith angle dependence of the $\nu_\mu$ events \cite{superk}
suggest oscillations of $\nu_\mu$ into $\nu_\tau$ with the
parameters $\sin^2 2\theta>0.7$ and $1.5\times 10^{-3}<\Delta m^2
<1.5\times 10^{-2}$ eV$^2$ \cite{superk}. This is in agreement
with previously reported results on the atmospheric anomaly
by Kamiokande \cite{kamioka} and MACRO \cite{macro} and is
consistent with limits from other experiments, e.g., CHOOZ \cite{chooz}.
Confirmation of 
$\nu_\mu \rightarrow \nu_\tau$ oscillations and 
determination of neutrino mixing angles would be 
a crucial indication of the 
nature of physics beyond the Standard Model.  
The firmest 
confirmation of this hypothesis would be via 
detection of $\tau$ leptons produced by charged current 
interactions of $\nu_\tau$'s resulting from 
oscillations of 
$\nu_\mu$'s, which is 
extremely difficult with current neutrino experiments.

In this letter, we propose a study of 
$\nu_\mu \rightarrow \nu_\tau$ oscillations with 
extragalactic neutrinos.   
Large volume neutrino detectors and the prospect of astrophysical neutrino
sources put $\nu_\tau$ detection in the realm of possibility.
The large distances involved for astrophysical sources, on the
order of one to thousands of Megaparsecs, make the next generation of
neutrino experiments potentially sensitive to neutrino mass
differences as low as $\Delta m^2\sim 10^{-17}$ eV$^2$ \cite{halzen}.
Over such long baselines, half of the neutrinos arriving at the
earth would be $\nu_\tau$'s in oscillation scenarios, the other
half being $\nu_\mu$'s. By observing both $\nu_\mu$ and $\nu_\tau$
from extragalatic sources such as Gamma Ray Bursts (GRBs) \cite{waxman} and
Active Galactic Nuclei (AGN) \cite{stecker}, 
neutrino oscillation hypothesis would be confirmed and
models of these sources would be tested.

The effect of attenuation of the neutrino flux due to interactions
of neutrinos in the Earth is qualitatively different for 
$\nu_\mu$ and $\nu_\tau$. Muon neutrinos are absorbed
by charged current interactions, while tau neutrinos are regenerated
by tau decays. The Earth never becomes opaque to $\nu_\tau$, though
the effect of $\nu_\tau\rightarrow \tau\rightarrow \nu_\tau$
interaction and decay processes is to degrade the energy of the incident
$\nu_\tau$. The identical spectra of $\nu_\mu$ and $\nu_\tau$ incident on the
Earth emerge after passage through the Earth with distinctly
different spectra.
The preferential penetration of $\nu_\tau$ through
the Earth is of great importance for high energy neutrino
telescopes such as AMANDA, NESTOR and ANTARES.

We consider $\nu_\mu$ and $\nu_\tau$ propagation
through the Earth using a similar procedure to the one outlined
for $\nu_\mu$ in Ref. \cite{Nauper}. We show that the energy spectrum
of the $\nu_\tau$ becomes enhanced at low energy, providing a distinctive
signature for its detection. The degree of enhancement depends on
the initial neutrino flux. We consider initial fluxes $F_\nu^0\sim
E^{-n}$ for $n=1,2,3.6$, a GRB flux \cite{waxman} and an AGN flux \cite{stecker}.
We solve the coupled transport equations for lepton and neutrino
fluxes as indicated below.

Let $F_{\nu_{\tau}}(E,X)$ and $ F_\tau(E,X)$ be the differential energy
spectrum 
of tau neutrinos and tau
respectively at a column depth $X$ in the medium defined by
\[
X = \int_0^L\rho(L')dL',
\]
where $\rho(L)$ is the density of the medium at a distance $L$ from
the boundary measured along the neutrino beam path. Then, one can derive 
the following cascade equation for neutrinos as,
\begin{eqnarray}
\nonumber
& &\frac{\partial F_{\nu_{\tau}}(E,X)}{\partial X} =
- \frac{F_{\nu_{\tau}}(E,X)}
	{\lambda_{\nu_{\tau}}(E)}
+ \int_E^\infty dE_y
\left[\frac{F_{\nu_{\tau}}(E_y,X)}{\lambda_{\nu_{\tau}}
(E_y)}\right]
	{\frac{dn}{dE}}({\nu_{\tau}}N\rightarrow {\nu_{\tau}}X; E_y,E)
\end{eqnarray}
\begin{eqnarray}
\nonumber
+ \int_E^\infty dE_y
\left[\frac{F_{\tau}(E_y,X)}{\rho_{\tau}^{dec}(E_y)}\right]
	{\frac{dn}{dE}}({\tau}\rightarrow {\nu_{\tau}}X; E_y,E)
\end{eqnarray}
\begin{eqnarray}
+\int_E^\infty dE_y
\left[\frac{F_{\tau}(E_y,X)}{\lambda_{\tau}(E_y)}\right]
	{\frac{dn}{dE}}({\tau}N\rightarrow {\nu_{\tau}}X; E_y,E)	
\end{eqnarray}
and for taus as,
\newpage
\begin{eqnarray}
\nonumber
\frac{\partial F_\tau(E,X)}{\partial X} =
- \frac{F_\tau(E,X)}{\lambda_\tau(E)}
	- \frac{F_\tau(E,X)}{\rho_\tau^{dec}(E,X,\theta)}
\end{eqnarray}
\begin{eqnarray}
+ \int_E^\infty dE_y
\left[\frac{F_{\nu_{\tau}}(E_y,X)}{\lambda_{\nu_{\tau}}
	(E_y)}\right]{\frac{dn}{dE}}({\nu_{\tau}}N\rightarrow {\tau}X;
E_y,E) .
\end{eqnarray}

The first term in Eq. (1) is a loss due to the neutrino 
interactions, the second is the regeneration term due 
to the neutral current, the third term is a contribution due 
to the tau decay and the last term is the contribution due to tau
interactions.  

In Eq. (2), the first term is a loss due to tau interactions, 
the second term is a loss due to the tau decay, 
while the last term is a contribution from neutrino 
charged current interactions.  
As a practical matter, tau decays are more important than
tau interactions at the energies considered here, though interactions 
become more important at higher energies \cite{seckel}.
We neglect the tau interaction terms in Eqs. (1) and (2) in what follows. 

Here $\lambda(E)$ is the interaction length and 
$\rho_{\tau}^{dec}(E,X,\theta)$ 
is the decay length for tau. They are defined as,
\[
\frac{1}{\lambda_\nu(E)}=\sum_T N_T\sigma_{\nu T}^{{tot}}(E),
\]

\[\rho_{\tau}^{dec}(E,X,\theta) = \gamma c \zeta_{\tau}\varrho(X,\theta)
\]
where $N_T$ is the number of scatterers $T$ in 1 g of the medium,
$\sigma^{{tot}}_{{\nu}T}(E)$ is the total cross section for
the ${\nu}T$  interactions and the sum is over all scatterer types
($T = N,e,\ldots$), $\zeta_\tau$ is the mean lifetime of 
tau and $\varrho$ is the density of matter in the earth.
Scatterings of neutrinos and taus with nucleons ($N$) are most important,
so we approximate
\[
\frac{1}{\lambda_\nu(E)}\simeq N_0\sigma_{\nu N}^{{tot}}(E),
\]
where $N_0$ is Avogadro's number. We use the CTEQ5 parton distribution
functions to evaluate neutrino cross sections \cite{cteq5}.
We have previously calculated charged and neutral current 
energy distributions, $dn/dE$, and the total cross section, 
$\sigma^{{tot}}_{{\nu}T}(E)$ taking into account recent improvements in 
our knowledge of 
the small-x behavior of the structure functions \cite{gqrs98}.  

To simplify the solution to the
equation for the tau flux, we approximate the $X$ and $\theta$
dependent density of the earth by the average of the density along
the column depth of angle $\theta$:
\[
\varrho(X,\theta)\simeq \varrho^{avg}(\theta)\ .
\]
Following Ref. \cite{Nauper},
let us define the effective absorption length $\Lambda_\nu(E,X)$ by
\begin{equation}
F_\nu(E,X) = F_\nu^0(E)\exp\left[-\frac{X}{\Lambda_\nu(E,X)}\right].
\end{equation}
\newpage
It is convenient to define
\begin{equation}
\Lambda_\nu(E,X) = \frac{\lambda_\nu(E)}{1-Z_{\nu}(E,X)}
\end{equation}
where $Z_{\nu}(E,X)$ is a positive function (we will call it $Z$ factor
in analogy with the hadronic cascade theory) which contains the
complete information about neutrino interaction and
regeneration in matter.

Assuming that there is no significant contribution to the neutrino
flux from decaying particles (as would be the case for muon
neutrinos)
using the above equation we can find an implicit equation for $Z$
from 
the transport equation  \cite{Nauper}, 
\begin{equation}
Z_{\nu}(E,X) = \int_0^1\eta_\nu(y,E)\Phi_\nu^{nc}(y,E)
\left[\frac{1-e^{-XD_\nu(E,E_y,X)}}
                  {XD_\nu(E,E_y,X)}\right] dy,
\end{equation}
where 
\[D_\nu(E,E_y,X)=\frac{1}{\Lambda_\nu(E_y,X)} - 
\frac{1}{\Lambda_\nu(E,X)}\]
\[\eta_\nu(y,E) = \frac{F_\nu^0(E_y)}{F_\nu^0(E)(1-y)}\]
\[\frac{d\sigma_{\nu N\rightarrow\nu X}(y,E_y)}{dy} =
\Phi_\nu^{nc}(y,E)\sigma_{\nu N}^{{tot}}(E)\]
and $d\sigma_{\nu N\rightarrow\nu X}(y,E)/dy$ is the
differential cross section for the inclusive reaction
$\nu N\rightarrow\nu X$ (with $E_{y}$ the incoming neutrino energy and
$y$ the fraction of energy lost) and $E_y \equiv E/(1-y)$.

Naumov and Perrone \cite{Nauper} have shown that by iteratively
evaluating Eq. (5), starting with $Z^{(0)}=0$, the solution 
for muon neutrinos quickly
converges for a wide range of starting fluxes.

By a similar procedure, the coupled differential equations 
for tau neutrinos including
tau production and decay can be iteratively solved. The tau flux
generated by charged current interactions including the loss term due to
its decay is
\begin{eqnarray}
\nonumber
{\frac{F_\tau(E,X)}{F_\nu^0(E)}}=
\exp\left[-\frac{X}{\rho_\tau^{dec}(E,\theta)}
\right]\int_0^X\int_0^1{\frac{\Phi_\nu^{cc}(y,E)}{\lambda_{\nu}(E)}}
\eta_\nu(y,E)
\end{eqnarray}
\begin{eqnarray}
\times
\exp\left[-\frac{X'}{\Lambda_\nu(E_y,X')}\right]
\exp\left[\frac{X'}{\rho_\tau^{dec}(E,\theta)}\right]dX^{'}dy \ .
\end{eqnarray}
The $Z$ factor for the tau neutrino flux is then
\begin{equation} Z = Z_{\nu} + Z_{\tau}
\end{equation}
where $Z_\nu$ is given by Eq. (5) and
\begin{eqnarray}
Z_{\tau} =  \left[\frac{1}{X}\right]
\int_0^X\int_0^1{\frac{\lambda_\nu(E)}{\rho_{\tau}^{dec}(E,\theta)}}
\Phi_\nu^{dec}(y,E)
\eta_\nu(y,E)\exp\left[-\frac{X'}{\Lambda_\nu(E_y,X')}\right]
{\frac{F_{\tau}(E_y,X^{'})}{F_{\nu}^0(E_y)}} dX^{'} dy.
\end{eqnarray}

We include decay modes in $\Phi_\nu^{dec}(y,E)$ as in
Ref. \cite{pasq} and a constant energy distribution for the remaining 
branching fraction not included there.
In Eqs. (5) and (8), the $Z$ factors implicit in $\Lambda_\nu$ are
$Z=Z_\nu+Z_\tau$.

In the iterative solution of the equation for $Z$, one has the option
of picking the initial value $Z^{(0)}$. We have chosen the $X$ and $E$
dependent solution to the cascade equation for the
$\nu_\mu$ flux, namely the solution to Eq. (5).

To demonstrate the importance of regeneration of tau neutrinos from
tau decays, we evaluate the tau neutrino flux for several input
neutrino spectra and compare to the attenuated $\nu_\mu$ flux.
For the incoming neutrino spectrum we use \cite{Nauper} 
\begin{equation}
F_\nu^0(E) = K\left(\frac{E_0}{E}\right)^{n}
                       \phi\left(\frac{E}{E_{\rm{cut}}}\right),
\end{equation}
where $K$, $n$, $E_0$ and $E_{\rm{cut}}$ are
parameters and $\phi(t)$ is a function equal to 0 at $t\geq1$ and 1
at $t \ll 1$. We use 
$\phi(t)=1/\left[1+\tan\left(\pi t/2\right)\right]$ ($t < 1$) and
$E_{\rm{cut}} = 3\times10^{10}$~GeV and  $ E_0 = 1$ PeV.
For $n=1$, we introduce a smooth cutoff by multiplying Eq. (9) by a factor
$\left(1+{E_0}/{E}\right)^{-2}$ and by setting $E_0=100$ PeV.
To evaluate the depth as a function of nadir angle, the density profile of
the earth described in  Ref. \cite{gqrs98} is used.

In Fig. 1 we show the nadir angle dependence of the ratios of the
fluxes calculated via Eqs. (1-8) to the
input flux $F_\nu^0(E)$. All fluxes are evaluated as a function of
nadir angle, at the $X$ value for the surface of the earth and 
for 
energies $10^4$ GeV, $10^5$ GeV and $10^6$ GeV for $\nu_\mu$ and $\nu_\tau$ 
assuming  
$F_\nu^0(E)$ given by Eq. (9).  
For an incoming flux, $n=1$, we find that 
the $\nu_\tau$ flux is enhanced (relative to the incoming $\nu_\tau$ flux) 
for all nadir angles for $E_{\nu_{\tau}}=
10^4$ GeV and $10^5$ GeV, and for $\theta > 20^\circ$ for $E_{\nu_{\tau}}=
10^6$ GeV.  The peak of the enhancement gets shifted toward the higher 
nadir angles as the energy increases.  This is due to the fact that 
high energy $\nu_\tau$ can remain high energy if the column depth
is small, i.e. for large nadir angles.  In case of the 
steeper incoming flux, $n=2$, we find that $\nu_\tau$'s are less attenuated 
than the $\nu_\mu$'s, and the expected enhancement at low energy is not 
evident due to the steepness of the flux.  For even steeper flux, 
$n=3.6$, the difference 
between $\nu_\tau$ and $\nu_\mu$ flux is very small.

In Fig. 2 we show the energy dependence of the 
the ratio of fluxes for nadir angles 
$\theta=0$, 
$\theta=30^\circ$ and 
$\theta=60^\circ$ for 
$\nu_\mu$ and $\nu_\tau$ with 
$F_\nu^0(E)$ given by Eq. (10).  For small nadir angles, $\theta=0$ and $30^\circ$
and 
$F_\nu^0(E)\sim 1/E$ we find that enhancement of tau neutrinos is in the 
energy range of $10^2$ GeV and $10^5$ GeV, while for 
$\theta=60^\circ$, 
the enhancement extends up to $10^6$ GeV.  In contrast 
the $\nu_\mu$ flux is attenuated for all the nadir angles.  When the 
incoming flux is steeper, $n=2$, the $\nu_\tau$ flux appears to be attenuated 
at high energies, although less than the $\nu_\mu$ flux.    For $n=3.6$, 
the energy dependence of these two fluxes is very similar, they are both 
reduced at high energies, and the effect is stronger for smaller nadir 
angle, since in this case the column depth is larger and there are more 
charged current interactions possible.

In case of the AGN quasar model \cite{stecker}, 
we find that the $\nu_\tau$ flux is a factor of $2$ to $2.5$ times larger than 
the input flux, for nadir angle, $\theta=0$.  This is shown in 
Fig. 3.  For larger angles, the effect is smaller.  Detection of AGN neutrinos 
would be optimal for small nadir angles and for $\nu_\tau$ with energy of 
$10^2$ GeV to $10^4$ GeV.

We also present results for the $\nu_\tau$ flux for the case of GRB fireball 
model \cite{waxman}.  We find that due to the steepness of the input 
flux for $E_{\nu_{\tau}}>100$ TeV, 
the $\nu_\tau$ flux is enhanced only by about $10-27\%$, 
depending on the energy and nadir angle.  We show the energy spectrum of the
ratio of $\nu_\tau$ flux to the input flux $F_\nu^0$ in Fig. 4.  

We expect that the next generation of neutrino telescopes will be 
able to detect the high energy neutrinos from AGN and GRBs.  
We have previously shown that 
most of the extragalactic neutrino fluxes 
exceed the atmospheric neutrino 
background for 
neutrino
energy greater than $\sim 10$ TeV which may enable the detection of the 
extragalactic neutrinos \cite{gqrs98}.  
A search for high-energy $\nu_\tau$ appearance would look for
enhanced muon rates.

The enhanced rates of muons come from muonic decays of the $\tau$
produced by $\nu_\tau$ charged current interactions in or near
the instrumented detector volume. The muon rates would be
enhanced at low energy ($\sim 10-100$ TeV) and for small nadir
angles.
In the case of an AGN quasar model, we find that the $\nu_\tau$
flux enhancement is very 
distinct, while for a GRB fireball model, the effect is only 
$10-27\%$.  This is due to the fact that GRB input $\nu_\tau$ flux 
is much steeper, thus there are not many high-energy $\nu_\tau$'s that 
would contribute to the enhancement at low energy.  We have also 
shown that the low energy pile up is significant only for 
incoming fluxes that are less steep than $1/E^2$ in the high energy 
region.  For an incoming flux which is proportional to $1/E$, we find 
the angular distribution of $\nu_\tau$'s to be significantly different 
than for the $\nu_\mu$'s.
   
We have proposed a novel way of detecting 
appearance of extragalactic high-energy $\nu_\tau$ 
by measuring the angular and energy distribution of muons with 
energy above 10 TeV.  This would give 
an experimental signature
of 
$\nu_\mu \rightarrow \nu_\tau$ oscillations with 
$\Delta  m^2$ as low as $10^{-17}$ eV$^2$ .


\vskip 0.1true in

\leftline{Acknowledgements}  

We are grateful to F. Halzen and D. Seckel for stimulating
discussions.  The work of S.I. and I.S. has been supported in part
by the DOE under Contract
DE-FG02-95ER40906.  The work of M.H.R. has been supported in
part by
National Science Foundation Grant No.
PHY-9802403.
M.H.R and I.S.  thank
Aspen Center for Physics for its hospitality while
part of this work was completed.  
Upon completion of this work, we
received a conference paper by Bottai and Becattini who had obtained 
similar results using a Monte Carlo calculation for a $1/E$ 
flux \cite{bb}.

\begin{figure}[!hbt]
\rule{0.0cm}{7.0cm}\\
\epsfxsize=15cm
\epsfbox[0 0 4096 4096]{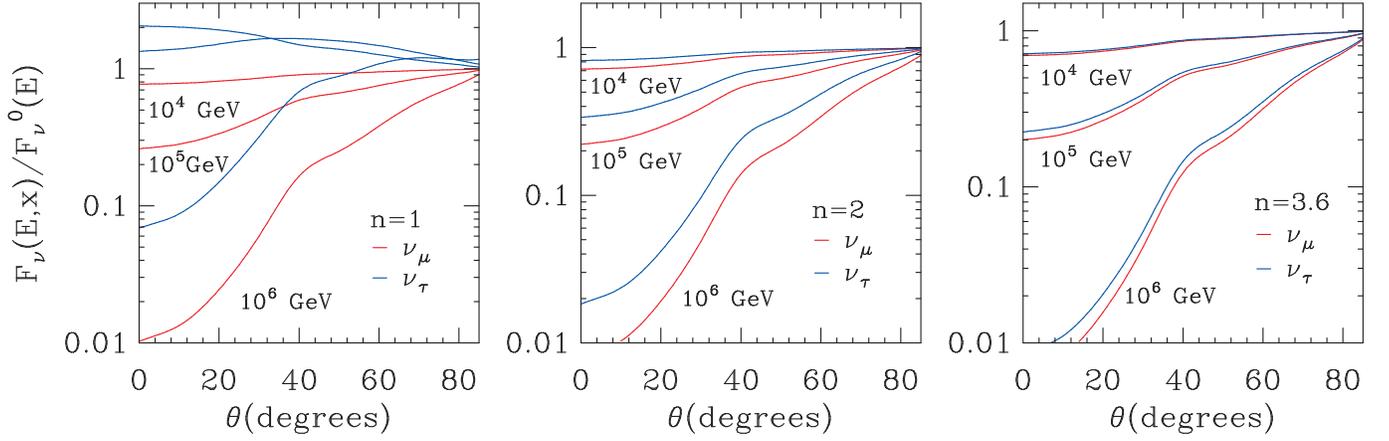}
\medskip
\vskip -2.0true in
\rule{0.0cm}{0.0cm}\vspace{-6.0cm}\\
\caption{ \normalsize{The nadir angle dependence of the ratio of fluxes for energies
$10^4$ GeV, $10^5$ GeV and $10^6$ GeV for the $\nu_\mu$ and $\nu_\tau$ assuming
$F_\nu^0(E)\sim E^{-n}$ with $n = 1,2$ and $3.6$.}
\label{tau}}
\end{figure}

\begin{figure}[!hbt]
\rule{0.0cm}{3.0cm}\\
\epsfxsize=15cm
\epsfbox[0 0 4096 4096]{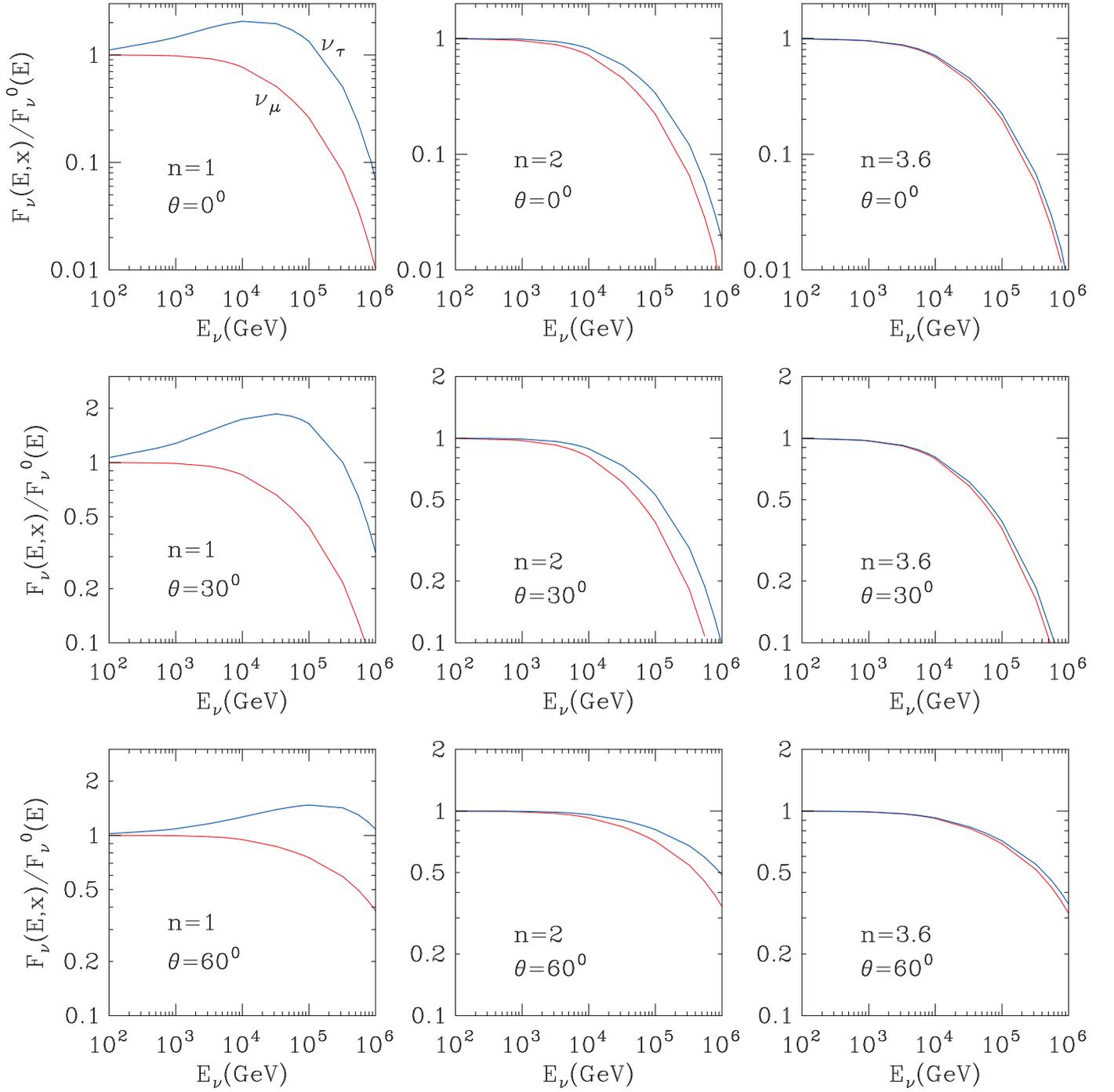}
\medskip
\rule{0.0cm}{1.0cm}\\
\medskip
\vskip 0.5true in
\caption{\normalsize{
The energy dependence of 
the ratio of fluxes for nadir angles 
$\theta=0$, 
$\theta=30^0$ and 
$\theta=90^0$ for 
$\nu_\mu$ and $\nu_\tau$ assuming 
$F_\nu^0(E) \sim E^{-n}$ with $n=1,2$ and $3.6$.}  
\label{atten}}
\end{figure}
\newpage

\begin{figure}[!hbt]
\rule{0.0cm}{7.0cm}\\
\epsfxsize=15cm
\epsfbox[0 0 4096 4096]{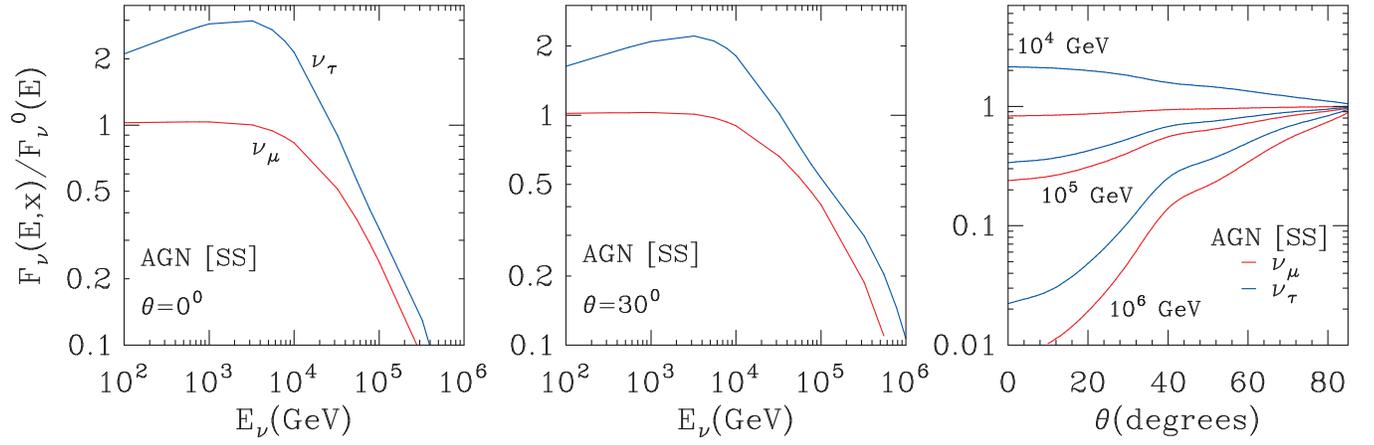}
\medskip
\rule{0.0cm}{0.1cm}\vspace{-6.0cm}\\
\medskip
\vskip -2.0true in
\caption{\normalsize{The energy dependence of the ratio of fluxes 
for a Stecker-Salamon AGN model [7].}}
\end{figure}
\newpage

\begin{figure}[!hbt]
\rule{0.0cm}{7.0cm}\\
\epsfxsize=15cm
\epsfbox[0 0 4096 4096]{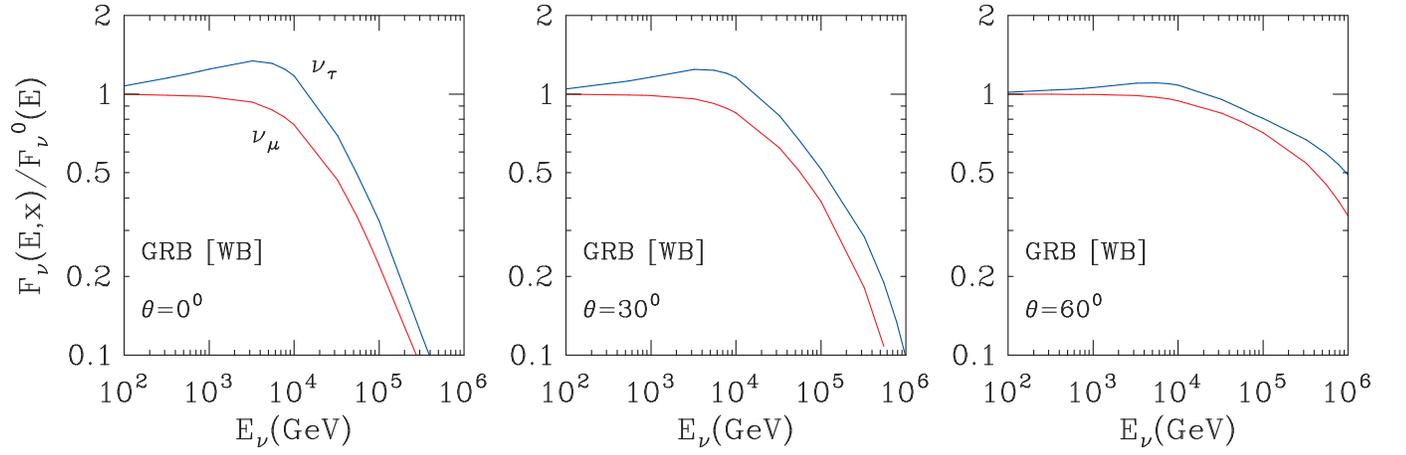}
\medskip
\rule{0.0cm}{0.1cm}\vspace{-6.0cm}\\
\vskip -2.0true in
\caption{\normalsize{The energy dependence of the ratio of the fluxes for a 
Waxman-Bahcall GRB model [6].}} 
\end{figure}
\end{document}
\bye